\documentclass[aps,pra,twocolumn,superscriptaddress]{revtex4-1}
\pdfoutput=1

\usepackage{amsmath}
\usepackage{amssymb}

\usepackage{xcolor}

\usepackage{setspace}
\usepackage{bm}
\usepackage{graphicx}
\usepackage{hyperref}
\usepackage[caption=false]{subfig}

\begin{document}


\title{An Autonomous Stabilizer for Incompressible Photon Fluids and Solids}
\author{Ruichao Ma}
\affiliation{James Franck Institute and the Department of Physics at the University of Chicago}
\author{Clai Owens}
\affiliation{James Franck Institute and the Department of Physics at the University of Chicago}
\author{Andrew Houck}
\affiliation{Department of Electrical Engineering at Princeton University}
\author{David I. Schuster}
\affiliation{James Franck Institute and the Department of Physics at the University of Chicago}
\author{Jonathan Simon}
\affiliation{James Franck Institute and the Department of Physics at the University of Chicago}
\date{\today}

\begin{abstract}
We suggest a simple approach to populate photonic quantum materials at non-zero chemical potential and near-zero temperature. Taking inspiration from forced evaporation in cold-atom experiments, the essential ingredients for our low-entropy thermal reservoir are (a) inter-particle interactions, and (b) energy-dependent loss. The resulting thermal reservoir may then be coupled to a broad class of Hamiltonian systems to produce low-entropy quantum phases. We present an idealized picture of such a reservoir, deriving the scaling of reservoir entropy with system parameters, and then propose several practical implementations using only standard circuit quantum electrodynamics tools, and extract the fundamental performance limits. Finally, we explore, both analytically and numerically, the coupling of such a thermalizer to the paradigmatic Bose-Hubbard chain, where we employ it to stabilize an $n=1$ Mott phase. In this case, the performance is limited by the interplay of dynamically arrested thermalization of the Mott insulator and finite heat capacity of the thermalizer, characterized by its repumping rate. This work explores a new approach to preparation of quantum phases of strongly interacting photons, and provides a potential route to topologically protected phases that are difficult to reach through adiabatic evolution.
\end{abstract}

\maketitle

\section{Introduction}
\label{sec:Intro}
Building synthetic materials relies upon the ability to engineer a desired many-body Hamiltonian, and a way to populate that Hamiltonian with particles at low temperature. With the advent of Rydberg EIT \cite{pritchard2010cooperative,peyronel2012quantum} and circuit QED \cite{wallraff2004strong,schuster2007resolving}, it is now possible to engineer strong interactions between individual long-lived photons  \cite{devoret2013superconducting}, making photonics an exciting place to begin to engineer quantum materials. It has become clear that photonic platforms are uniquely suited to the task, offering exquisite control of single-particle dynamics: These efforts have led to realizations of photonic kagome \cite{TunnelNumbers2012} and honeycomb \cite{polini2013artificial} lattices, synthetic magnetic fields for photons \cite{wang2009observation,hafezi2013imaging,rechtsman2013photonic,ningyuan2015time,schine2016synthetic}, and numerous proposals to explore strongly correlated quantum phases in photonic systems, using the unique input-output capabilities provided by an optical platform \cite{umucalilar2013many,grusdt2014topological}.

An upcoming challenge in photonic systems is populating the system Hamiltonian with particles, such that they reside in a low-entropy many-body state. Recent works have demonstrated bath engineering ~\cite{shankar2013autonomously,mirrahimi2014dynamically} as a viable approach for stabilizing small entangled photon states. To thermalize photonic manybody phases, several proposals suggest creating a chemical potential through parametric driving \cite{kapit2014,hafezi2015}. Here we present and thoroughly explore an alternative which is directly applicable to stabilization of incompressible phases, and does not face the heating mechanisms expected to be present in Floquet models \cite{RoschFloquet2015}.

Our approach is based upon the development of a narrowband, continuously replenished photon source akin to those demonstrated for individual quantum dots in \cite{Koski2015,leek2009using}: we prepare this source  by creating a population inversion (near-complete occupation of $n=1$ state) of a single non-linear resonator via either (1) a $0\rightarrow 2$ drive, and Purcell-enhanced $2\rightarrow 1$ decay, thereby stabilizing the 1-photon state; or (2) photon-photon collisions in a Wannier-Stark ladder which drives one photon into a stabilized resonator and the other into a resonator providing Purcell-enhanced decay. In both approaches, the key is to combine Purcell-enhanced loss with strong interactions to provide a channel for shedding entropy.

Global control of electron density is easily achieved in the solid-state, where charge-conservation imposes a strong constraint on the total number of electrons. Because photons are uncharged, the density of photons in a synthetic material is harder to control. We rely upon a discontinuity in the chemical potential at a particular photon number to stabilize our photon density; this may be understood as requiring that the target phase be incompressible.

It is worth briefly contrasting this approach with that in ultracold atomic quantum gases: laser cool an atomic gas, transferring atomic entropy to a scattered optical field, and then remove the remaining entropy through evaporative cooling: atomic collisions leading to loss of high-energy atoms, and subsequent rethermalization. This procedure prepares a Bose-Einstein condensate, a low-entropy phase of matter which may be smoothly converted into many other phases \cite{Greiner2002,Bakr2010,Simon2011} by adiabatically varying the system Hamiltonian, thereby crossing quantum phase transitions. This approach is ideal for cold-atoms, where the dynamics are slow, the tools to create and manipulate the Hamiltonian are global, and low entropy BECs are readily available as a starting point. By contrast, the strength of photonic systems lies in local manipulation and readout \cite{shanks2013scanning,ma2016hamiltonian} of the many-body state, while real-time tuning of the Hamiltonian is more challenging to achieve because dynamics are $\sim 6$ orders of magnitude faster than for a typical atomic quantum gas in an optical lattice.

Adiabatic preparation requires tuning through a quantum phase transition \cite{Greiner2002,Bakr2010,Simon2011}, where the many-body gap closes, while thermalization \cite{mathy2012enlarging} or algorithmic cooling \cite{Bakr2011} into an interacting phase only necessitate competition with the many-body gap away from critical points. Consequently, for a constant product of sample lifetime ($\tau$) and interaction- ($U$) or tunneling- ($J$) energy, the system entropy in a cooled system can be much lower than that of corresponding (a) spectroscopically \cite{CarusottoFQH2012,HafeziFQHScaling2013} or (b) adiabatically~\cite{zurek2005dynamics} prepared systems, where defects are induced by vanishing wave-function overlaps or small energy gaps, respectively.

In Sec. \ref{sec:thermideas} we motivate the need for tools to autonomously stabilize photonic states by describing the challenge of optically pumping a qubit. In Sec. \ref{sec:SimpleModel} we explore approaches to creating a photonic thermalizer using coupled non-linear resonators and engineered dissipation; in Sec. \ref{sec:MBCoupling} we analyze the performance of the thermalizer when it is actually coupled to an interacting lattice model, demonstrating that the approach is effective for stabilizing a Bose-Hubbard model near an $n=1$ Mott phase.

\section{Stabilization Concepts}
\label{sec:thermideas}
It is straightforward to populate a strongly interacting photonic lattice with photons; driving a single lattice site with a laser pulse, RF tone, or even random noise, will suffice. The challenge is that none of these approaches stabilizes the system near the many-body ground state at finite photon number. This may be understood by attempting to stabilize a single lattice site with a single photon, which we will now explore in the circuit quantum electrodynamics paradigm, where the photonic lattice is an array of capacitively coupled qubits. In this language, our objective is to stabilize a single qubit in its first excited state.  Coherent driving will induce the qubit to go through a repeated process of Rabi oscillation and decay, and at long times will be in a statistical mixture of ground and excited states, with a maximal excited state probability $P_e\leq\frac{1}{2}$. To stabilize the qubit in the excited state ($P_e\approx 1$) in steady state, then, requires a more sophisticated scheme. One might imagine the following classical feedback procedure: $\pi$-pulse the qubit and continuously monitor its state, applying another $\pi$-pulse whenever the it decays. We analyze an \emph{autonomous} version of this process which is not limited by detection path quantum efficiencies, which is a simplified version of prior bath engineering proposals~\cite{pastawski2011quantum,verstraete2009quantum,mirrahimi2014dynamically,kraus2008preparation,diehl2008quantum}.

The essential element for stabilization of any system in a particular state is a channel into which entropy may be shed. A classical harmonic oscillator, for example, stabilizes at zero amplitude only if it has damping-- otherwise it continues to coherently oscillate forever. More broadly, the entropy of a system may be shed into a classical measurement channel, as in the scheme described above, or, taking examples from existing synthetic materials, it can be shed into an emitted light field, particle loss channel, or phonon bath, in the cases of laser- cooling, evaporative- cooling, and exciton-polariton condensation \cite{byrnes2014exciton} respectively. In what follows, we take specific inspiration from evaporative cooling: entropy is pumped out of a system when particles collide and one achieves sufficient energy to leave the trap, while the other's energy is reduced. We describe a way for a qubit to ``decay'' into its excited state, shedding its entropy into an evaporated photon by using an engineered bath. We then demonstrate that by coupling this qubit to an interacting many-site lattice the whole system will be stabilized near its many-body ground state.

\section{Simple Model of a Narrowband Stabilizer}
\label{sec:SimpleModel}
We need to create a single lattice site with near-continuous single photon occupation (so-called ``population inversion'') that rapidly repumps itself to single-photon occupancy whenever the photon in it leaves, either due to particle loss from finite resonator lifetime or tunneling into a tunnel-coupled many-body system ~\cite{kapit2014}.

\begin{figure}
\includegraphics[width=\columnwidth]{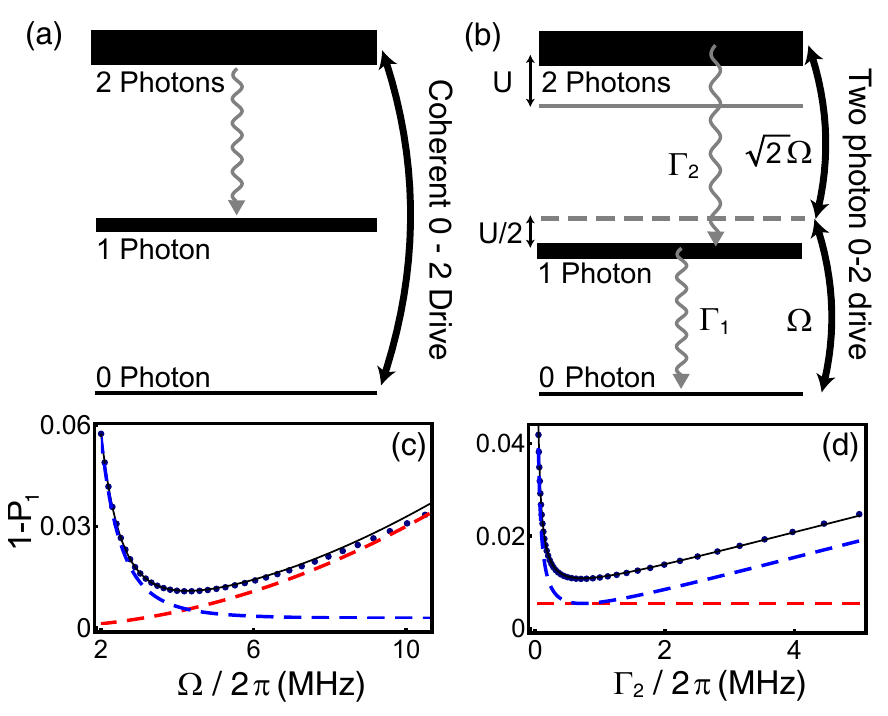}
\caption{\label{Figure:MostIdealThermalizer} \textbf{Idealized Model of Continuously Inverted Qubit}. \textbf{(a)} To prepare a three-level qubit in its first excited state, it can be continuously resonantly excited $0\rightarrow 2$, and allowed to rapidly decay (via an engineered loss channel) into a long-lived one-photon state. \textbf{(b)} More realistic model including the requisite anharmonicity $U$ to prevent accidental resonant excitation $0\rightarrow 1$; the optimal performance is $1-P_1=2\sqrt{6\Gamma_1/U}-6\Gamma_1/U$, for single photon loss rate $\Gamma_1$. In essence, too little drive $\Omega$ allows the system to spend excessive of time in the 0-photon state after a decay event and too much drive produces coherent admixture of zero- and two- photon states in the single photon state. \textbf{(c)} \& \textbf{(d)} compare the analytic model (solid curves) to master equation numerics as drive Rabi frequency $\Omega$ and two-photon loss rate $\Gamma_2$ are varied, respectively. For these simulations we study state-of-the-art qubits with $\Gamma_1\approx 2\pi\times 1$ kHz, and $U\approx 2\pi\times200$ MHz \cite{schuster2007resolving,NarrowQubit2012}. All other parameters chosen to be their analytical optima described in the text. The dashed curves indicate the contributions to thermalizer error coming from off-resonant admixture of 0- and 2- photon states (red), or single-particle loss (blue).}
\end{figure}

We propose to create the inversion through a variant of optical pumping, depicted qualitatively in Fig.~\ref{Figure:MostIdealThermalizer}a, where the $2$-photon state is made short-lived, and the $1$-photon state long-lived. The idea is then to drive a qubit directly from the 0-photon state to the 2-photon state, from which it will rapidly decay into the 1-photon state; we thus need the 2-photon lifetime to be very short compared with the 1-photon lifetime. Before we suggest specific implementation of the $2^{nd}$-photon loss channel, we compute the performance of a simplified model with freely adjustable $2^{nd}$-photon loss (not two-photon loss; only the second of the two photons is rapidly removed).

Because the $0 \rightarrow 2$ photon transition is not directly allowed, we drive it through a two-photon transition with the $1$ photon intermediate state off-resonant due to the qubit anharmonicity $U$ (see Fig.~\ref{Figure:MostIdealThermalizer}b). With a one-photon Rabi coupling $\Omega$, and an $n$-photon loss rate $\Gamma_n$, one can write the probability of single photon occupation $P_1$ as (in the low-infidelity limit):

\begin{align}
1-P_1 &=&12\frac{\Omega^2}{U^2} + \left[1+\frac{\Gamma_2}{\Gamma_1}\left(2+\frac{{\Gamma_2}^2}{32\left(\frac{\Omega^2}{U}\right)^2}\right)^{-1} \right]^{-1}\nonumber\\
\end{align}

The first term comes from off-resonant admixture of $0$- and $2$- photon states into the stabilized state, and the second term from the competition between the single-particle loss in the $1$ photon state, $\Gamma_1$, and the (saturated) pumping rate into the $1$-photon state through the two-photon incoherent coupling $0 \rightarrow 2 \rightarrow 1$. Put simply: too little driving and the system spends a lot of time in $0$ due to the one-body $1\rightarrow0$ decay; too much driving, and the coherent admixture of the $0$ and $2$ photon states becomes large.

The one-photon probability is maximized for $\Gamma_2^{opt}=8\frac{(\Omega^{opt})^2}{U}$, $\Omega^{opt}=\left(\frac{U^3\Gamma_1}{24}\right)^{1/4}\sqrt{1-\sqrt{\frac{6\Gamma_1}{U}}}\approx \left(\frac{U^3\Gamma_1}{24}\right)^{1/4}$, yielding $\left<1-P_1\right>_{optimal}=2\sqrt{\frac{6\Gamma_1}{U}}-\frac{6\Gamma_1}{U}$; we do not optimize over $U$ or $\Gamma_1$, as these parameters are set by the experimental state of the art. In the low-temperature limit, this $P_1$ yields (for chemical potential $\mu=U/2$) a qubit temperature of $k_B T\approx \frac{U}{\log{\frac{U}{24\Gamma_1}}}$, and an entropy of $\frac{S}{k_B}\approx 2 e^{-\frac{U}{2k_B T}}\left(1+\frac{U}{2 k_B T}\right)$ (see Appendix \ref{sec:SIthermStatMech}).

For state-of-the-art parameters \cite{schuster2007resolving,NarrowQubit2012}, $U\approx 2\pi\times200$ MHz, $\Gamma_1\approx 2\pi\times 1$ kHz, performance is optimized for $\Gamma_2\approx 2\pi\times 730$ kHz, $\Omega\approx 2\pi\times4.3$ MHz. One achieves $\left<1-P_1\right>_{optimal}\approx 1.1\times 10^{-2}$, and corresponding temperature $k_B T\approx 0.1 \times U$ and entropy $S\approx 0.1 \times k_B$. Figure~\ref{Figure:MostIdealThermalizer}c,d compares this simple analytic theory with the results of a numerically solved master equation model (see appendices \ref{sec:SInumerics} and \ref{sec:SInumericsB}), describing the steady state probability of unit photon occupancy, and demonstrating quantitative agreement.

\begin{figure}
\includegraphics[width=\columnwidth]{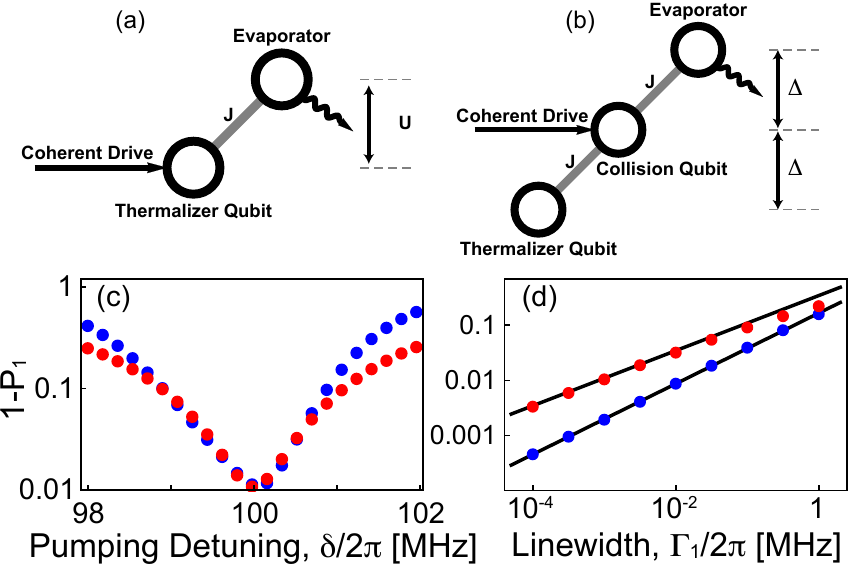}
\caption{\label{Figure:ThermalizerDesigns} \textbf{Thermalizer Designs}. \textbf{(a)} To implement particle number dependent loss in the ``thermalizer'' qubit, it is tunnel coupled to a lossy site (the ``evaporator'') which is detuned by the onsite interaction energy U. Thus while it is energetically forbidden for a first photon to leave the ``thermalizer'' qubit, the second can leave at no energy cost through the ``evaporator''. The system is excited from the vacuum state to the two-photon state via an off-resonant two-photon transition through the one-photon state, from which it rapidly decays to the long-lived one-photon state. \textbf{(b)} A higher-performance three-qubit approach employs a central qubit (the ``interaction'' qubit) where photon pairs may resonantly collide, with one driven into the ``thermalizer'' qubit and the other driven into the ``evaporator'' qubit, from which it is immediately lost, thereby preventing the collision process from reversing itself. This design relies upon strong interactions in the ``thermalizer'' qubit to suppress double excitations. \textbf{(c)} Comparison of master-equation simulations of (red points) idealized model with two-photon loss-rate $\Gamma_2$, and (blue points) realistic model of two-photon loss implemented as in (a), with $J\approx 2\pi\times730$ kHz, $\kappa\approx2\pi\times5$ MHz. Both curves are plotted versus the pump detuning $\delta$, which is optimized at $\delta=U/2$ where the two curves agree, as anticipated. \textbf{(d)} Comparison of two (red)- and three (blue)- qubit thermalizers, versus the qubit lifetime $\Gamma_1$, for fixed qubit anharmonicity $U$; optimized over all other parameters. The master-equation numerics for the two-qubit thermalizer reveal that the occupation error $1-P_1$, scales as $(\frac{\Gamma_1}{U})^{1/2}$, while the error of the three-qubit thermalizer scales as  $\left(\frac{\Gamma_1}{U}\right)^{2/3}$.}
\end{figure}

The second-photon loss $\Gamma_2$ is the key to this technique, and may be introduced through tunnel-coupling to a lossy qubit/resonator (the ``evaporator''), tuned to resonance only with the second photon in the primary qubit (See Fig.~\ref{Figure:ThermalizerDesigns}a). For an evaporator with linewidth $\kappa$, and a tunnel-coupling strength $J$, it is straightforward to show that $\Gamma_2\approx\frac{2J^2}{\kappa}$, while $\Gamma_1$ is slightly increased due to off-resonant coupling to the thermalizer, $~\frac{J^2}{U^2}\kappa$. This increase in $\Gamma_1$ may be kept below $\Gamma_1$ itself by choosing $\kappa\leq\sqrt{2\frac{\Gamma_1}{\Gamma_2}} U$, $J\leq\sqrt{U\sqrt{\frac{\Gamma_1\Gamma_2}{2}}}$, or by using a Purcell filter~\cite{houck2008controlling,reed2010fast}. Figure~\ref{Figure:ThermalizerDesigns}c compares the performance of such a two-site thermalizer to the idealized thermalizer analyzed above, demonstrating good agreement between the two.

An alternate approach, shown in Fig.~\ref{Figure:ThermalizerDesigns}b and similar to \cite{holland2015single,Souquet2016}, employs three degrees of freedom to achieve better performance. A central (``collision'') qubit is driven with a coherent tone, and anharmonicity-induced photon-photon collisions split photon pairs, driving one to the upper (``evaporator'') qubit/resonator, and one to the lower (``thermalizer'') qubit. The photon in the ``evaporator'' is quickly lost, leaving only the photon in the ``thermalizer'' qubit, which cannot Rabi-flop back into the ``collision'' qubit due to conservation of energy. A second photon is precluded from scattering into the ``thermalizer'' qubit due to an anharmonicity-induced photon blockade. While analytics for this more sophisticated model are prohibitively complex, a numerical optimization of its performance (see Fig.~\ref{Figure:ThermalizerDesigns}d) indicates $1-P_1\approx4.4\times\left(\frac{\Gamma_1}{U}\right)^{0.64}\sim\left(\frac{\Gamma_1}{U}\right)^{2/3}$, which is more favorable than the $\left(\frac{\Gamma_1}{U}\right)^{1/2}$ scaling of the two-qubit thermalizer.

\begin{figure}
\includegraphics[width=0.9\columnwidth]{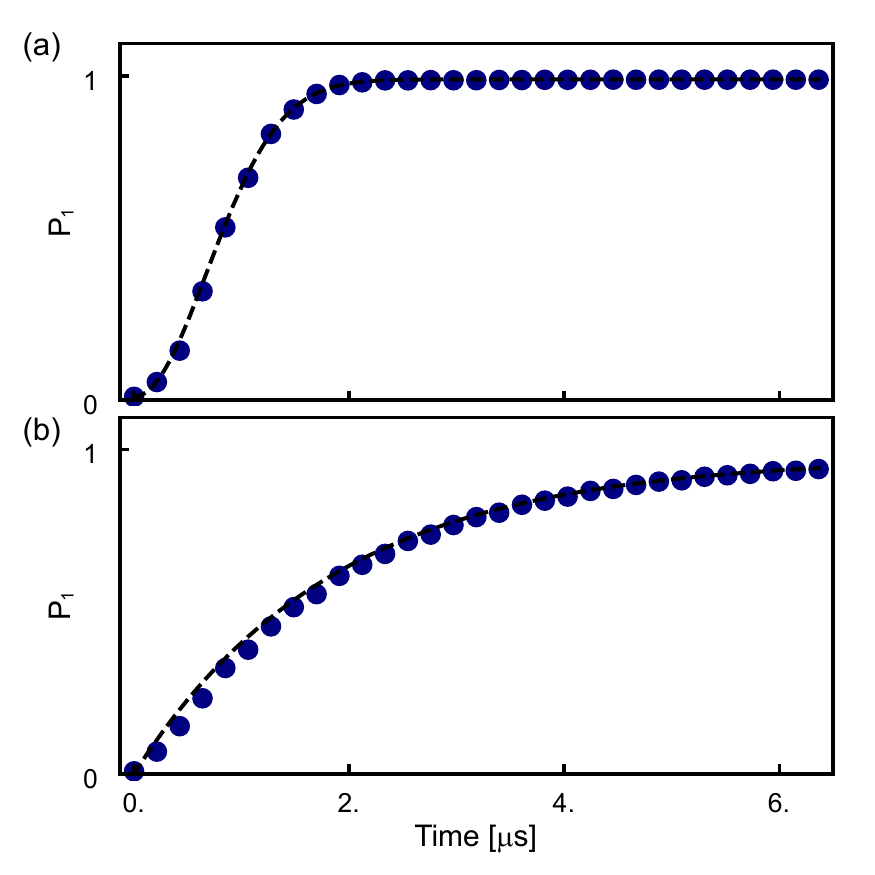}
\caption{\label{Figure:ThermalizerDynamics} \textbf{Refilling dynamics of the two-site thermalizer.} Subsequent to a photon loss event (either through tunneling of the photon into an attached manybody system, or finite qubit $T_1$), the resonator refills through states $|0\rangle \rightarrow |2\rangle \rightarrow |1\rangle$. We consider a qubit with an anharmonicity $U=2\pi\times200$ MHz, and a $T_1$-limited linewidth $\Gamma_1=2\pi\times1$ kHz. (a) Refilling dynamics under the conditions which provide the optimal $P_1$, as described in the text. The evolution towards $P_1\approx 1$ is non-exponential, as the $0\rightarrow 2$ excitation process is saturated. The theory curve (dashed) is a gaussian empirical model of width (time to 1/e) 0.9$\times\tau$, with $\tau^{-1}\approx\sqrt{\frac{\Gamma_1 U}{6}}$. (b) For $\sim4.3$ times larger $\Gamma_2$, the system operates in an over-damped regime, and refills exponentially towards $P_1\approx 1$.}
\end{figure}

In preparation for exploring the coupling of this reservoir to a many-body system, we numerically investigate the repumping dynamics of the thermalizer after its photon is lost, either through spontaneous decay or tunneling into the many-body system. While this may be understood formally in terms of the smallest non-zero eigenvalue of the Louivillian; we take a simple, physical approach here: Figure~\ref{Figure:ThermalizerDynamics} shows the temporal dynamics of $P_1$ after such a photon-loss event in a two-site thermalizer. In (a), the dynamics occur under conditions that minimize thermalizer temperature, leading to critical damping, and repumping with a $e^{-1}$ time-constant $\tau\approx R^{-1}$, for a repumping rate $R\approx 0.9 \sqrt{\frac{\Gamma_1 U}{6}}$ (implied by equation \ref{Figure:ThermalizerDynamics}); (b) depicts the over-damped, purely exponential dynamics which occur for increased 2$^{nd}$-photon loss. In the presence of coupling to a many-particle system the optimal parameters change because the thermalizer is more often depleted and thus must repump faster to efficiently stabilize the system. To this end, in Appendix \ref{sec:SIrepFixed} we show that the optimal infidelity of a thermalizer with repumping rate $R_t$ is given by (optimizing over the drive strength $\Omega$ and the $2^{nd}$-photon loss rate $\Gamma_2$):

\begin{equation}
\label{eq:thermfidelity}
1-P_1=6\frac{R_t}{U}+\frac{\Gamma_1}{\Gamma_1+R_t}
\end{equation}

In what follows, we adopt a simplified model of the thermalizer, treating it as a device which exponentially decays towards single occupancy with a chosen repumping rate $R_t$, and infidelity $1-P_1$ given by Eqn. \ref{eq:thermfidelity}.

\section{Coupling to a Many-Body System}
\label{sec:MBCoupling}
Employing the thermalizer to stabilize a many-body system in a particular phase requires that this phase be incompressible \emph{from the perspective of adding particles} rather than varying the volume.

\begin{figure}
\includegraphics[width=\columnwidth]{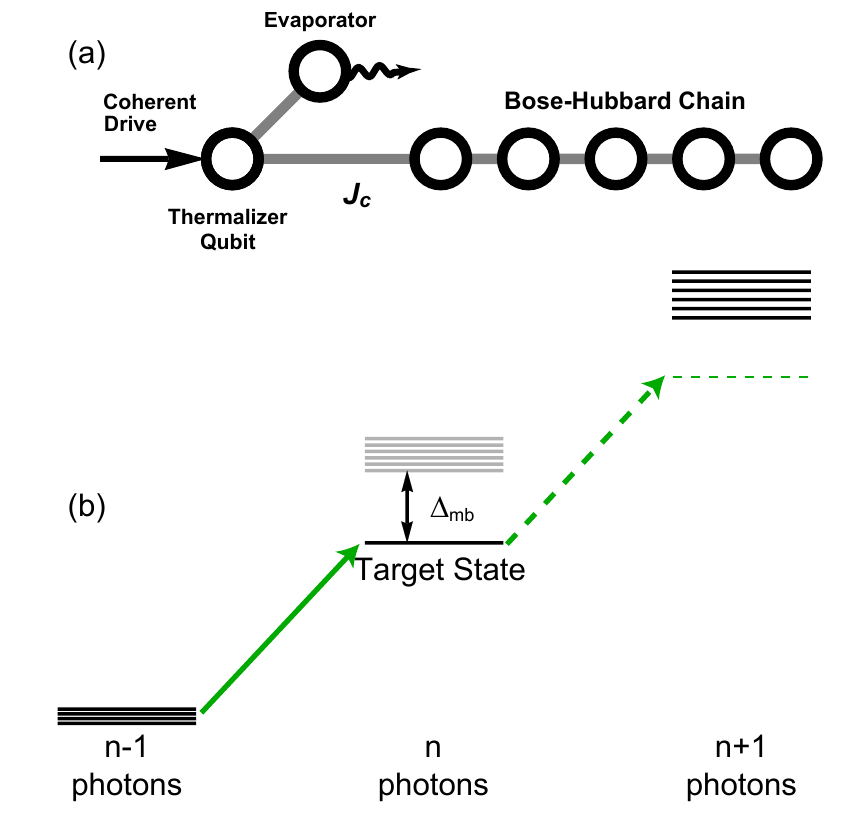}
\caption{\label{Figure:CouplingThermalizer}  \textbf{Coupling the thermalizer to a manybody system}. (a) The thermalizer proposed in the text may be employed to stabilize an incompressible phase of a manybody system. This is achieved by tunnel-coupling the thermalizer to the system (with a strength $J_c$), and allowing the two to come into equilibrium. (b) Operational principle. The thermalizer relies upon the difference between the energy cost to add and remove a particle (what we call \emph{incompressibility}). This is because the thermalizer stabilizes the particle number by only providing particles of certain energies. As such, the energy to remove a particle from- (add a hole to- ) the manybody ground state (shown in green) must be spectroscopically resolved from both the energy to remove a particle from a many-body excited state (with a gap $\Delta_{\mathrm{mb}}$) and the energy to add a particle to the many-body ground state.}
\end{figure}

In coupling a thermalizer to a many-particle system, it is essential to match the spectral width of the thermalizer to the hole-spectrum of the system (see Fig.~\ref{Figure:CouplingThermalizer}a). To understand this, consider the state of the system to be near an incompressible phase into which we would like to stabilize it (see Fig.~\ref{Figure:CouplingThermalizer}b). If the system is already in the phase, we need to ensure that we do not inject additional particles. On the other hand, if the system is a single particle short of being in the appropriate state (because a photon recently decayed), the thermalizer must be able to inject a photon of the appropriate energy to re-excite the system into the incompressible ground state, but not into an excited state with the same number of particles. It is thus crucial that the hole- and particle- spectra be non-overlapping energetically: this imposes an additional constraint on the system to be populated and is what is meant by ``incompressible'', rather than the more standard definition $\frac{\partial V}{\partial P}=0$ at fixed particle number. The two definitions are equivalent if the energy per particle depends only upon the particle density (see Appendix \ref{sec:SIcompressibility}).

Once this incompressibility criterion is met, the next question is \emph{how efficiently} a thermalizer can refill defects in the many-body state-- essentially a question of Franck-Condon overlaps. The idea is to compute the spectrum of hole-like defects, and compute how efficiently each is repumped by the thermalizer; if such defects are excessively localized, a single thermalizer will not suffice to repump them, and thermalizers will be necessary at each site \cite{kapit2014}. For mobile defects, a single thermalizer suffices:

To demonstrate this, we now explore the stabilization of an $n=1$ Mott phase as a paradigmatic example of thermalizer performance. This is a particularly simple case to consider because hole-like excitations live near the bare-resonator energy (henceforth $E\approx 0$), while all particle-like excitations live near $E=U$; each band has a width $\sim J$ the tunneling energy, and in the Mott phase $U\gg J$, providing clear spectral separation between particle- and hole- bands. If a hole tunnels into the thermalizer site (with a tunneling rate $J_c$), it is refilled (``damped out'') at a rate $R$, as derived in the preceding section. One can thus build a simple model to investigate the refilling rate of an isolated hole by examining the spectrum of a single ``particle'' (hole) hopping in a 1D tight-binding lattice at rate $J$, with a single lossy site at the end with imaginary energy $R$ (loss of the hole corresponds to refilling into the target Mott phase), into which the hole may tunnel with a rate $J_c$.

It is apparent that $J_c$ controls the Franck-Condon overlap of the various quasi-hole states with the thermalizer;  $J_c=\sqrt{2}\times J$ is found to provide a quasi-hole overlap (and thus refilling rate) that is independent of hole quasi-momentum/energy in the limit $\frac{R}{J}\rightarrow 0$ (see Appendix \ref{sec:SIJcProof}). In Fig.~\ref{Figure:HoleRepumping}, we plot (for a chain of length $N=120$ sites), the refilling rate of a hole as a function of its energy. It is apparent that, save for a few states near $E=\pm 2J$, all hole states are refilled equally efficiently, at a rate $\approx\frac{R}{N}$. The final few hole states within about $0.048 \times \frac{R^2}{J}$ of $E=\pm 2J$ (those with quasi-momentum $|q|, |q-\pi|$, or $|q+\pi| \leq q_c\approx 0.22\times \frac{R}{J}$) are refilled substantially more slowly, at a rate $\approx \frac{R}{N}\times (\frac{q}{q_c})^2$ (for $|q|\leq q_c$). This is because hole-modes near the center and edges of the Brillioun zone have low group-velocity and are thus Zeno-suppressed from moving relative to the thermalizer; this leaves modes with extremely low refilling rate, along with neighboring modes with enhanced refilling.

\begin{figure}
  \includegraphics[width=0.9\columnwidth]{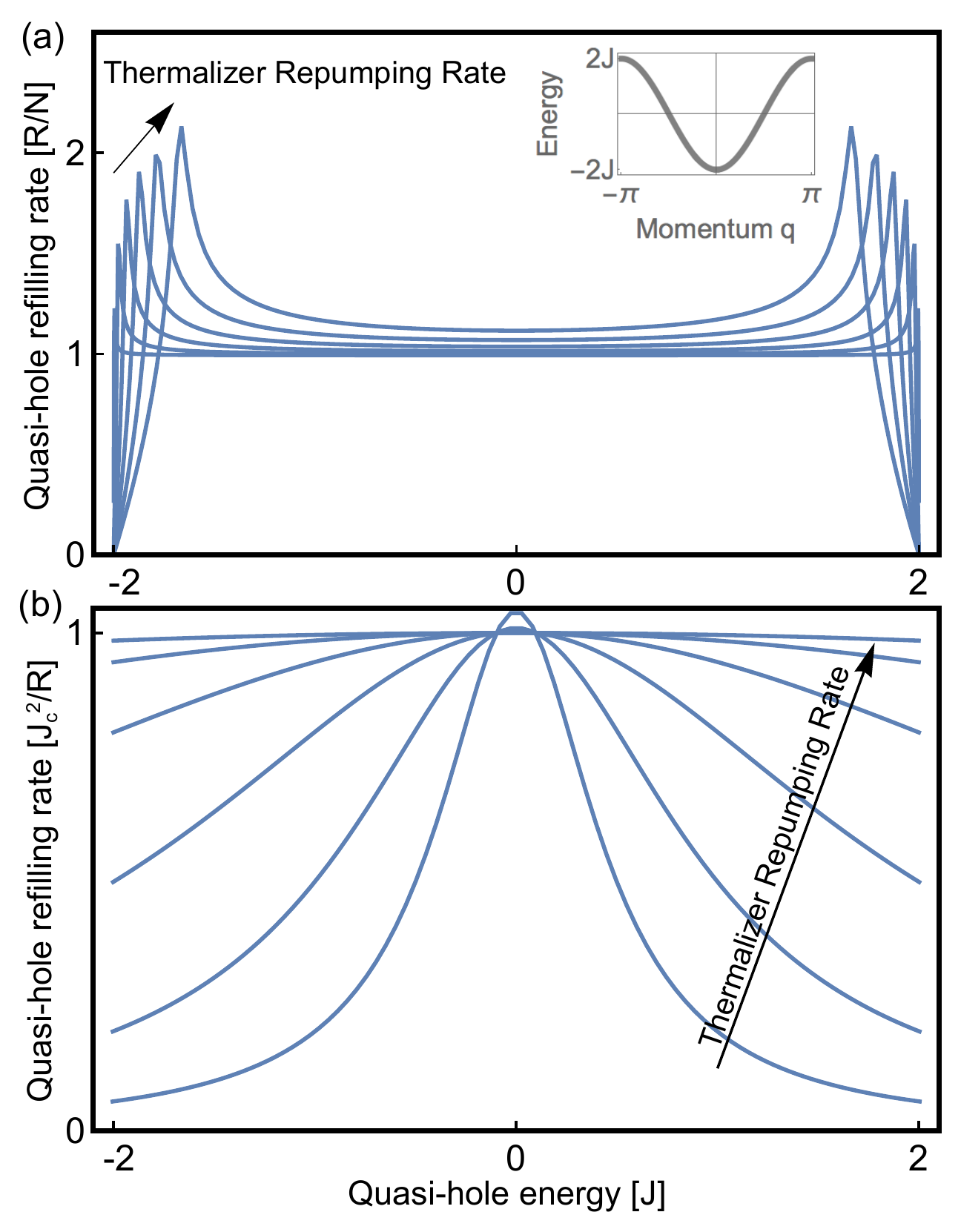}
\caption{\label{Figure:HoleRepumping} \textbf{Quasi-hole refilling dynamics}. The refilling rate of a quasi-hole in a 1D Mott Insulator is shown as a function of its energy, for repumping rates $R$ from $0.2J$ ... $2J$. The Mott phase lives in a 1D Hubbard-regime lattice coupled, in (a), to a thermalizer at one end. The energy of the quasi-hole reflects its quasi-momentum $q$ according to the relation $E=-2J\times\cos q$ (see Inset); as such, it is unsurprising that the lowest- and highest- energy quasi-holes are refilled inefficiently, as they exhibit low group velocity; equivalently their high density of states results in repumping-induced localization (from the Zeno effect) of some modes near the thermalizer (with faster refilling) in addition to those localized away from the thermalizer (with slower refilling). All refilling rates are normalized to $R/N$, the value expected in the low-repumping-rate limit, where quasi-holes at all quasi-momenta are refilled at the same rate. The theory shown is for a 120-site chain. In (b) a thermalizer is coupled directly to each site of the Bose-Hubbard chain, allowing for good Franck-Condon overlap with all quasi-holes states. The Lorentzian energy dependence of the refilling rate comes entirely from the detuning of the quasi-holes from the thermalizer, compared with the repumping rate. As such, we choose $R=J\times\left\{1,2,4,8,16\right\}$, for $J_c=0.1\times J$. Both (a) and (b) are computed using effective single-particle theories: we plot imaginary- vs. real- parts of the eigenvalues of a 1D tight binding model coupled to thermalizer sites whose repumping is modeled as an imaginary energy.}
\end{figure}

One can then build a simple model of the steady state defect probabilities in different quasi-hole modes as a competition between their mode-dependent refilling rate $r_q$, and their mode-independent creation rate $\Gamma_1$. The defect probability in quasi-hole mode with momentum $q$ is then given by $\epsilon_q=\frac{\Gamma_1}{\Gamma_1+r_q}$, where the repumping rate of mode $q$ is:
\[
    r_q= \frac{R}{N}\times
\begin{cases}
    1,& \text{if } |q|,|\pi-q|,|\pi+q|\geq q_c\\
    (\frac{q}{q_c})^2, & \text{if } |q|< q_c\\
    (\frac{\pi-q}{q_c})^2, & \text{if } |\pi-q|< q_c\\
    (\frac{\pi+q}{q_c})^2, & \text{if } |\pi+q|< q_c
\end{cases}
\]

where $q_c$ is the emperically-determined quasi-hole momentum cutoff defined above.

It is apparent that the performance of an ideal (error-less) thermalizer with repumping rate $R$ employed to stabilize a Bose-Hubbard chain near the $n=1$ Mott phase is a tradeoff between:
\begin{enumerate}
\itemsep-1em
\item low repumping rate, where photon decay at a rate $\Gamma_1$ competes with the repumping at a rate $R$ to limit the fidelity of the Mott phase; and
\item high repumping rate, where low-group velocity defects are poorly repumped.
\end{enumerate}

We can now compute the mean defect probability $\langle \epsilon \rangle=\frac{1}{2\pi}\int_{-\pi}^{\pi}\epsilon_q\mathrm{d}q$ by averaging over defects at different quasi-momenta. Performing this integral piece-wise yields:
\begin{align}
\label{eq:chainpumpfidelity}
\langle \epsilon \rangle&=&\left(1-\frac{2 q_c}{\pi}\right)\frac{1}{1+\frac{R}{N\Gamma_1}}+\frac{2 q_c}{\pi}\int_0^1\frac{\Gamma_1}{\Gamma_1+x^2 R/N}\mathrm{d}x\\
				        &=&\left(1-\frac{2 q_c}{\pi}\right)\frac{1}{1+\frac{R}{N\Gamma_1}}+\frac{2 q_c}{\pi}\sqrt\frac{N\Gamma_1}{R}\arctan\sqrt\frac{R}{N\Gamma_1}\\
				        &\leq&\frac{1}{1+\frac{R}{N\Gamma_1}}+q_c\sqrt\frac{N\Gamma_1}{R}
\end{align}

where the last line is worst-case performance. Noting that $q_c\equiv 0.22\times \frac{R}{J}$, we can optimize over R, with the result that (for $\Gamma_1\ll J,R$) $R_{optimal}\approx5.9\times(N \Gamma_1 J^2)^{1/3}$ and $\langle\epsilon\rangle\approx 0.7\times \left(\frac{N\Gamma_1}{J}\right)^{2/3}$. It is thus apparent that for fixed on-site loss $\Gamma_1$, it is favorable to maximize the tunneling rate $J$ to allow defects to leave the system as quickly as possible, increasing the thermalizer repumping rate accordingly.

\begin{figure}
\includegraphics[width=0.9\columnwidth]{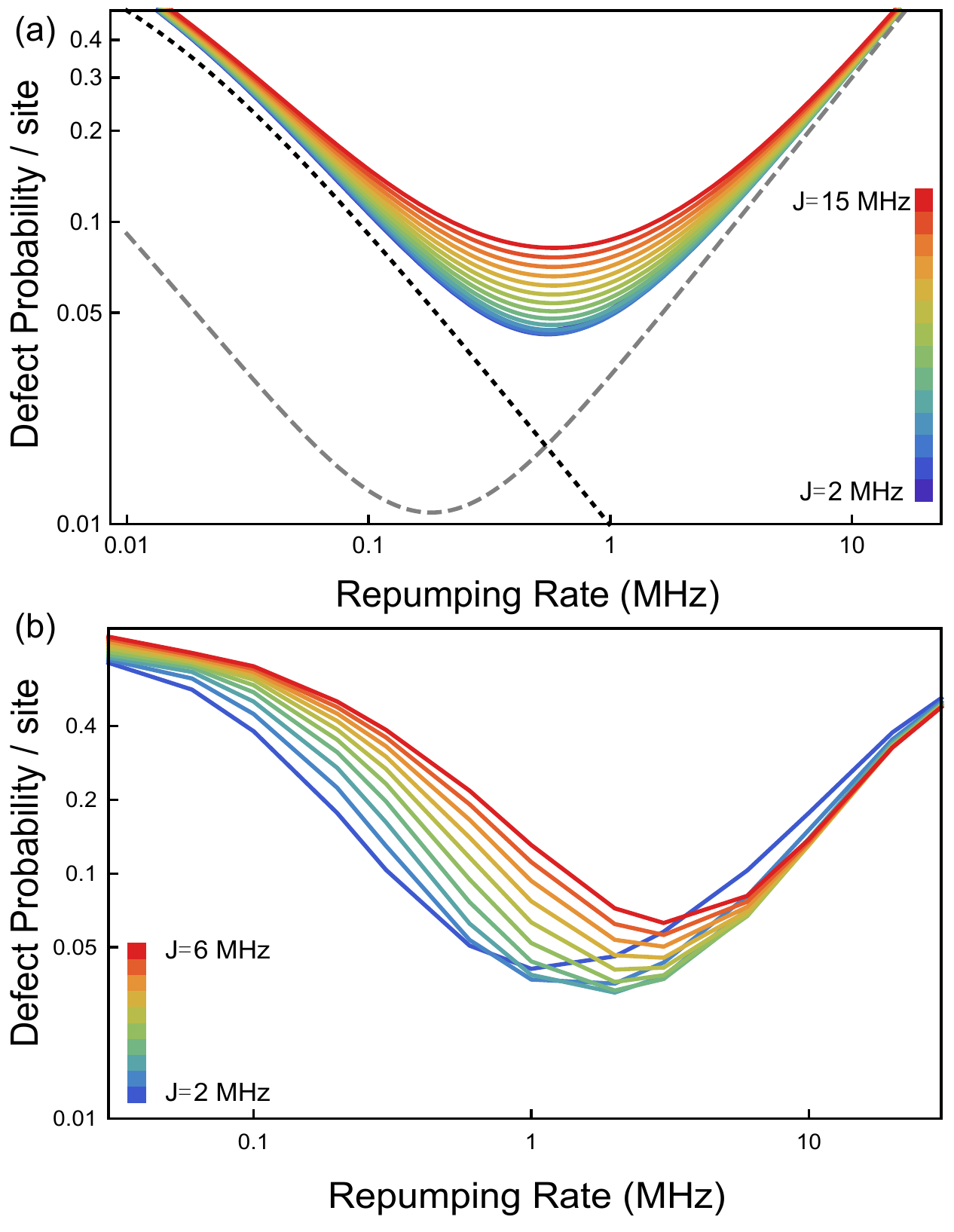}
\caption{\label{Figure:MIRepump} \textbf{Stabilizing a 1D Mott Phase}. In (a) we explore the behavior of an analytic model of defects in a 10-site Bose-Hubbard chain, stabilized with the two-site thermalizer described in the text. For photon-decay-rate $\Gamma_1=2\pi\times 1$ kHz and qubit anharmonicity $U=2\pi\times 200$ MHz, we plot the chain-averaged defect rate for various tunneling strengths and thermalizer repumping rates. While the performance is relatively insensitive to the choice of chain-tunneling-rate $J$, the optimum is at $J\approx 2\pi\times 6.5$ MHz, with an defect rate $\langle \epsilon \rangle\approx 0.04/$site. The observed performance is primarily limited by the tradeoff between insufficient cooling capacity of the thermalizer at low repumping rate (black dotted curve) and thermalizer errors at high repumping rate (gray dashed curve).  The inability of the thermalizer to repump defects at the highest and lowest quasi-momenta is not a limiting factor for the parameters explored here. The high infidelity of the chain at large $J$ comes from quantum-fluctuation-induced doublon-hole pairs. In (b) we employ a master-equation to explore a 3-site Bose-Hubbard chain coupled to an optimized 2-site thermalizer (modeled with effective $2^{nd}$ photon decay $\Gamma_2$). We find behaviour which qualitatively agrees analytic model, providing a slightly better optimal defect rate $\langle \epsilon \rangle\approx 0.03/$site as a result of the shorter chain.}
\end{figure}

In practice, other things limit the tunneling rate, including doublon-hole excitations of the Mott insulator and sensitivity of the realistic thermalizer to repumping rate (characterized by Eqn. \ref{eq:thermfidelity}): Fig. \ref{Figure:MIRepump}a shows the computed occupation infidelity of a 10-site Bose-Hubbard chain stabilized in the $n=1$ Mott phase, including defects from an imperfect thermalizer as in Eqn. \ref{eq:thermfidelity}, incomplete ability to refill defects in the chain as in Eqn. \ref{eq:chainpumpfidelity}, and coherent doublon-hole pairs due to non-zero $\frac{J}{U}$. Thus, for state-of-the-art parameters \cite{NarrowQubit2012,TunnelNumbers2012}, $\Gamma_1=2\pi\times 1$ kHz, and $J=2\pi\times 12$ MHz, a thermalizer with an anharmonicity of $U=2\pi\times200$ MHz can stabilize a $N=10$ site Mott insulator with a defect probability of $\langle \epsilon \rangle \approx 0.04/$site, when the repumping rate of the thermalizer is chosen to be $R \approx 2\pi\times 600$ kHz.

It is computationally challenging to perform a full master-equation simulation of a 10-site Bose-Hubbard chain coupled to such a thermalizer, so we have instead applied this formalism to a smaller 3-site chain coupled to a thermalizer, as shown in Fig.~\ref{Figure:MIRepump}b, for parameters similar to figure \ref{Figure:MIRepump}a. In this case, the optimal defect probability of $0.03/$site is achieved for a tunneling rate $J\approx2\pi\times 3.5$ MHz, and repumping rate $R\approx 2$ MHz, qualitatively validating the analytic approach explored in the preceding discussion. The necessity of higher repumping rate (and thus $\Gamma_2$) in Fig.~\ref{Figure:MIRepump}b relative to the analytic model in (a) likely arises at least in part from the tunnel-coupling of the $2^{nd}$ excited state of the thermalizer to the doublon-band of the chain. Note that some degree of doublon-repumping must be occurring to properly stabilize the Mott phase, and turns out to be induced by the second-photon loss intrinsic to the thermalizer (see Appendix~\ref{sec:SIrepumpingdoublons}).

Lower defect density could be achieved at the expense of greater experimental complexity by coupling a separate thermalizer to each site in the Bose-Hubbard chain. In this case the Franck-Condon overlap to the thermalizer is the same for all quasi-hole states, so the refilling rate of quasi-holes is determined entirely by their detuning from the thermalizer according to: $r_q\approx R\times \frac{J_c^2}{2J_c^2+8J^2\cos^2{q}+R^2}$. Here $R$ is the repumping rate of each thermalizer, $q$ is the quasi-momentum of the quasi-hole, $J$ is the tunneling rate in the Bose-Hubbard chain, and $J_c$ is the strength of the coupling of each site in the chain to its thermalizer. Note that as $J_c$ approaches and surpasses $R$, the thermalizers become part of the the many-body-system, rather than merely a device for stabilizing it. To achieve a nearly-uniform refilling rate for all quasi-holes, it is important to choose $R>2J$. This physics is explored in Fig.~\ref{Figure:HoleRepumping}b, where quasi-hole refilling rate is plotted as a function of quasi-hole energy. As anticipated, refilling rate becomes largely independent of quasi-hole energy once $R>2J,2J_c$.


\section{Outlook}
\label{sec:Outlook}
In this paper we propose a new approach to populating photonic Hamiltonians that is particularly well suited to those with gapped ground states. Harnessing the interplay of engineered dissipation and driven anharmonic oscillators, we develop a ``thermalizer'' which is autonomously stabilized in a state containing a single quantized excitation. This thermalizer may be attached to a Hubbard-type Hamiltonian system which it will then populate up to a gap in the particle-insertion spectrum, or equivalently a jump in the chemical potential. We demonstrate the efficacy of this approach by analytically and numerically coupling the ``thermalizer'' to a 1D Hubbard chain tuned to support a Mott phase. We explore how the thermalizer repopulates holes in the Mott phase of varying quasi-momenta, resulting in a comprehensive theory of the dynamics of quasi-hole refilling in a Mott insulator.

Looking forward, this thermalizer concept may be extended to more sophisticated models by tailoring the density of states of the thermalizer. This may be achieved by modulating either the frequency of thermalizer qubit, or its coupling strength to the manybody system. Such an approach would be beneficial for stabilization of n$=2$-and-higher Mott phases, as well as Laughlin-like topological ground states of flux-threaded 2D Hubbard lattices~\cite{anderson2016engineering}, and potentially even many-body localized states~\cite{serbyn2014interferometric}.

It is also possible to apply these ideas to cold atoms. Bilayer atomic quantum gas experiments \cite{preiss2015quantum} could be engineered such that one layer acts as a superfluid reservoir, coherently populating a site in the other layer which is itself engineered to act as a ``blockaded'' atomic reservoir \cite{Bakr2011,kantian2016dynamical} akin to what we have explored in this work. It would also be highly fruitful to apply these techniques to stabilize topological or crystalline phases of Rydberg polaritons \cite{ningyuan2016observation,sommer2015quantum}, where strong interactions \cite{peyronel2012quantum} may be combined with synthetic gauge fields in curved space \cite{schine2016synthetic,sommer2016engineering}.

Broadly, photonic systems now routinely achieve interaction-to-coherence ratios which compete with their atomic gas counterparts \cite{devoret2013superconducting}, and our approach connects these tools to synthetic materials, pointing the way to direct-cooling into quantum many-body phases of photonic systems.

\section{Acknowledgements}
\label{sec:Acknowledgements}
We would like to thank Tim Berkelbach, Aashish Clerk, Mohammad Hafezi, Ningyuan Jia, Nathan Schine, Aziza Suleymanzade and Jake Taylor for fruitful discussions. This work was supported by ARO grant W911NF-15-1-0397 for modeling of isolated thermalizers, and DOE grant DE-SC0010267 for exploration of hole-repumping. R.M. acknowledges support from the UChicago MRSEC Grant NSF-DMR-MRSEC 1420709.

\appendix
\section{Optimal Performance of a Thermalizer at Fixed Repumping Rate}
\label{sec:SIrepFixed}
Suppose we want to optimize the performance of the thermalizer subject to fixed refilling rate. That is: $R$ should be held at some value $R_t$.  As before, the single-excitation probability of the isolated thermalizer is given by:
\begin{align}
1-P_1 &=&12\frac{\Omega^2}{U^2} + \left[1+\frac{\Gamma_2}{\Gamma_1}\left(2+\frac{{\Gamma_2}^2}{32\left(\frac{\Omega^2}{U}\right)^2}\right)^{-1} \right]^{-1}\nonumber\\
\end{align}

but now we want to fix the refilling rate:
\begin{equation}
R_t =\Gamma_2\left(2+\frac{{\Gamma_2}^2}{32\left(\frac{\Omega^2}{U}\right)^2}\right)^{-1}\nonumber\\
\end{equation}

Optimizing subject to this constraint yields: $\Omega=\sqrt{\frac{R_t U}{2}}$; $\Gamma_2=4 R_t$; and $1-P_1=6\frac{R_t}{U}+\frac{\Gamma_1}{\Gamma_1+R_t}$.

\section{Achieving Uniform Refilling Rate of All Quasi-Holes in 1D Bose-Hubbard Chain}
\label{sec:SIJcProof}
In the text, we state that all quasi-holes refill at the same rate if the tunnel-coupling of the thermalizer to the Bose-Hubbard chain is given by $J_c=\sqrt{2}\times J$, where $J$ is the tunnel coupling in the chain itself. This is true in the limit that the repumping rate $R$ of the thermalizer site is much smaller than the tunneling rate of the chain, so we will here consider the limit of vanishing $\frac{R}{J}$, and prove constructively that all modes of the 1D tight-binding chain have equal probability in the thermalizer, assuming the coupling to the thermalizer is $\sqrt{2}$ times that of the chain itself:

Consider a uniform 1D tight-binding chain with tunneling rate $J$, and length $2M+1$, where $M$ is a positive integer (this chain does not have a thermalizer site which is more strongly coupled). The eigenmodes of this chain have energies $E_m=-2J\times\cos{\frac{\pi}{2}\frac{m}{M+1}}$, and mode functions $\psi_m(n)=\frac{1}{\sqrt{M+1}}\left\{\sin{},\cos{}\right\}\left(n\frac{\pi}{2}\frac{m}{M+1}\right)$ for $m$ $\{$even,odd$\}$ respectively, and sites indexed $n\in\left[-M,\ldots,M\right]$. We note that modes with odd $m$ are even about $n=0$, and vice-versa.

\emph{Now fold this chain in half, and merge site $n$ with site $-n$.} This may be achieved formally by strongly tunnel-coupling (with strength $J_{big}$) sites $n$ and $-n$, creating even- and odd- sub-manifolds separated in energy by $\sim J_{big}$. We now consider the spatially even manifold (that is, those with $m$ odd), whose eigenmodes $\theta_m(n)$ we already know: $\theta_m(n)=\frac{1}{\sqrt{2}}\left(\psi_m(n)+\psi_m(-n)\right)$, unless $n=0$, in which case $\theta_m(0)=\psi_m(0)$; the corresponding eigenvalues are also the same (up to the $J_{big}$ offset): $E_m=-2J\times\cos{\frac{\pi}{2}\frac{m}{M+1}}$.

The last step is to recover the tight binding model that this new chain obeys: a simple analysis reveals that all sites but site $n=0$ (of which there are $M$) are tunnel-coupled to their (properly normalized) neighbors with a tunneling rate $J$, while site $n=0$ is coupled to its neighbor with a tunneling rate $\sqrt{2}J$: this is the chain that we wanted to study! The wave-function overlap of each $m$-odd mode with site $n=0$ is $\theta_m(0)=\frac{1}{\sqrt{M+1}}\approx \frac{1}{\sqrt{M}}$ for $M\gg 1$, which completes the proof.

\section{Numerical Modeling of a Simplified Thermalizer}
\label{sec:SInumerics}
We model the thermalizer using a master equation, starting with a unitary Hamiltonian: 
\begin{equation}
H=\delta a^\dagger a+\frac{U}{2}a^\dagger a^\dagger a a + \Omega(a^\dagger+a)
\end{equation}
where $\delta$ is the pump-to-qubit detuning, $U$ is the qubit anharmonicity, and $\Omega$ is the pump Rabi frequency. We can parameterize the qubit anharmonicity with a single variable because we intend to operate almost exclusively in the singly-excited qubit state, so higher order contributions to the anharmonicity, which impact only to the third excited state and higher, are negligible.

To add dissipation to this model, we employ a master equation:
\begin{align}
\frac{\mathrm{d}\rho}{\mathrm{d}t}&=-i [H,\rho]+\frac{\Gamma_1}{2}\mathcal{L}[\rho,a]+\frac{\Gamma_2}{4}\mathcal{L}[\rho,a^\dagger a a]\nonumber\\
\mathcal{L}[\rho,C]&=\rho C^\dagger C+C^\dagger C\rho-2 C \rho C^\dagger
\end{align}

The first Lindblad term induces linear loss at a rate $R_{n\rightarrow n-1}=n \Gamma_1$, while the second Lindblad term induces nonlinear loss, at a rate $R_{n\rightarrow n-1}=\frac{n(n-1)}{2}\Gamma_2$; that is, it does not induce any loss for the $n=1$ state. We refer to this nonlinear Lindblad term as ``second photon loss'', because it removes the second photon but leaves the first; it should be compared to $\mathcal{L}[\rho, a a]$, which is a nonlinear Lindblad term that removes two photons at a time, and is typically referred to as ``two photon loss''.

For our numerics we solve for the steady state $\frac{\mathrm{d}\rho}{\mathrm{d}t}=0$, allowing up to 4 excitations in the system to ensure numerical convergence. For the three-site Bose-Hubbard chain we allow up to 6 excitations in the system.

\section{Numerical Modeling of a Realistic Thermalizer}
\label{sec:SInumericsB}
The most realistic models employed in this work model the nonlinear loss as it is experimentally implemented: via additional tunnel-coupled qubits and detuned resonators with their own linear loss terms. The resulting Hamiltonian takes the form:
\begin{align}
H&=\sum_j[(\Delta_j+\delta) a_j^\dagger a_j+\frac{U_j}{2}a_j^\dagger a_j^\dagger a_j a_j+t_j ({a_j}^\dagger a_{j+1}+h.c.)]\nonumber\\
&+\Omega({a_D}^\dagger+a_D)
\end{align}

Here $\Delta_j$ is the energy offset of the $j^{th}$ qubit; $U_j$ is the anharmonicity of the  $j^{th}$ qubit ($U_j=0$ when the $j^{th}$ qubit is in fact merely a lossy resonator); $t_j$ is the tunneling matrix element between the $j^{th}$  and  $j+1^{st}$ qubits; and $D$ is the index of the qubit which is driven. The dynamics then arise from the master equation:

\begin{align}
\frac{\mathrm{d}\rho}{\mathrm{d}t}=-i [H,\rho]+\sum_j\frac{\Gamma_j}{2}\mathcal{L}[\rho,a_j]
\end{align}

where $\Gamma_j$ is the linewidth of the $j^{th}$ qubit.

\section{Statistical Mechanics of the Single Qubit Thermalizer}
\label{sec:SIthermStatMech}
\begin{figure}
\includegraphics[width=0.95\columnwidth]{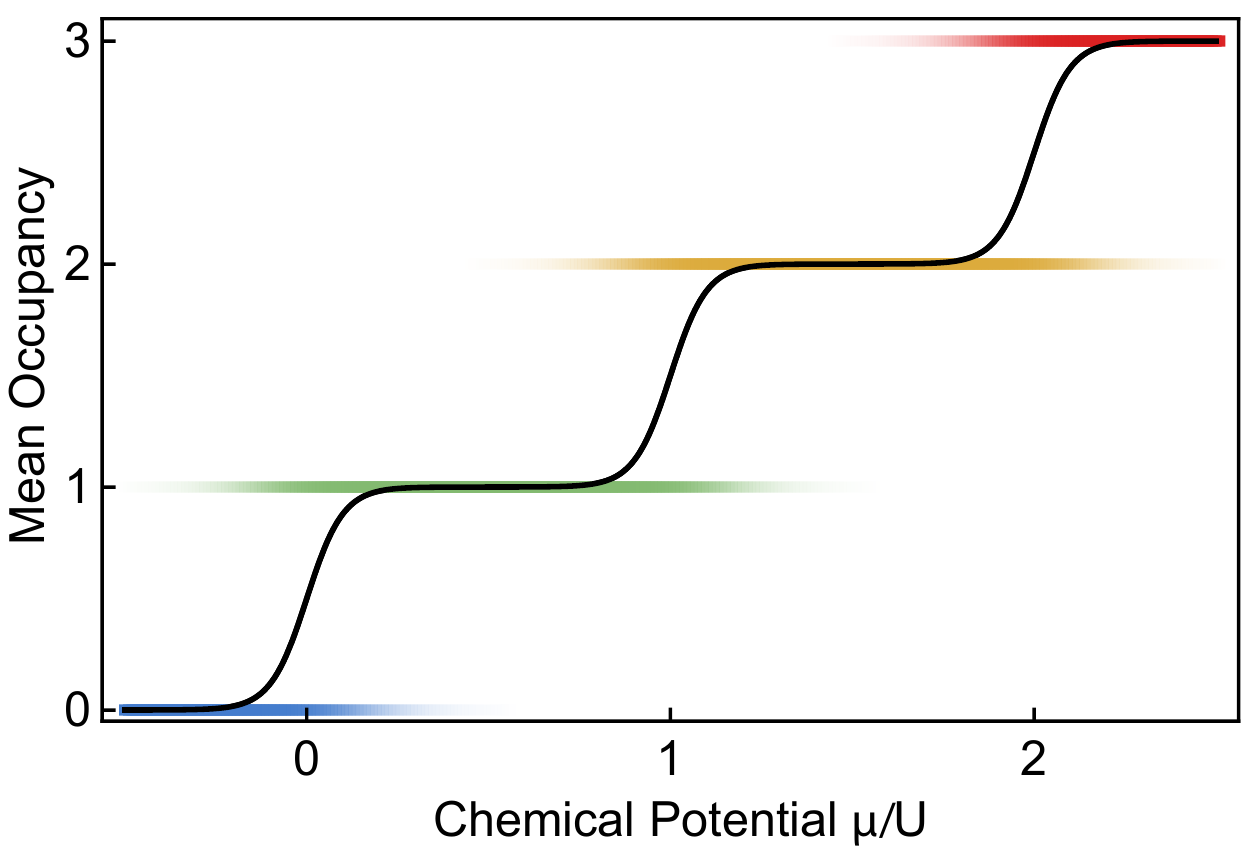}
\caption{\label{fig:SIchempotl}  \textbf{Understanding the Chemical Potential of a Generic, Isolated Thermalizer}. The mean occupancy $\langle n \rangle$ of an isolated, phenomenological thermalizer is computed, as a function of chemical potential $\mu$ in units of the thermalizer interaction energy $U$. The thermalizer is assumed to be in contact with a bath of temperature $k_B T= 0.1\times U$, and chemical potential $\mu$. As the chemical potential passes through multiples of the thermalizer intraction energy U, the occupancy $\langle n \rangle$  jumps in integer steps. The probability of occupancy $n$ is plotted as a colored horizontal line at height $n$, whose intensity reflects the probability.}
\end{figure}

\begin{figure}
\includegraphics[width=\columnwidth]{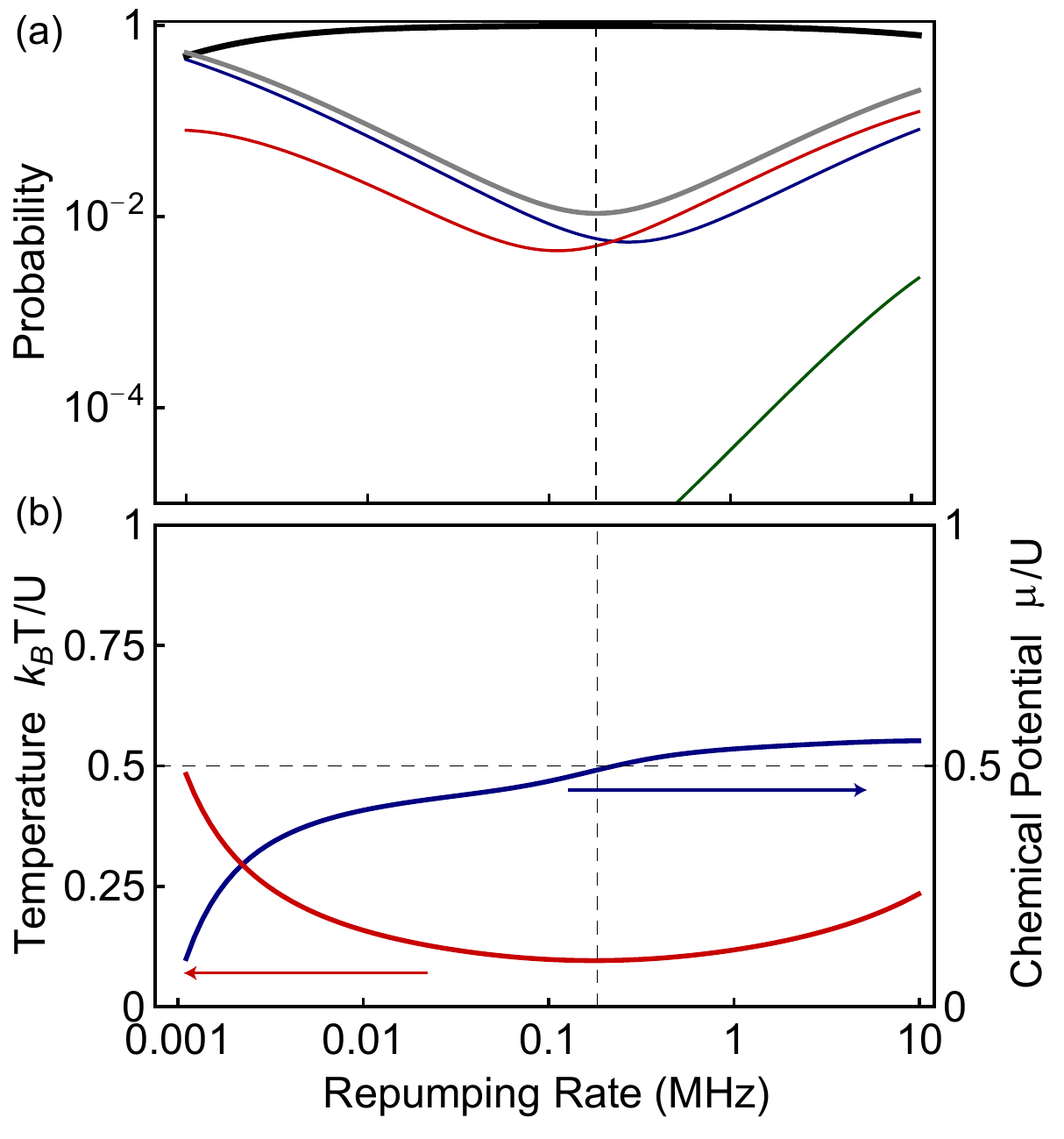}
\caption{\label{fig:SIchempotlNUMs}  \textbf{Quantifying a Realistic Thermalizer in terms of its Chemical Potential and Temperature}. (a) Full performance characterization of a two-site thermalizer, as a function of repumping rate, for the optimal qubit parameters in the text. (blue,black,red,green) are respectively the probability that the occupancy of the thermalizer is (0,1,2,3), with the gray curve representing the probability that the thermalizer is \emph{not} singly occupied. The vertical dashed line is at the optimal repumping rate (where $P_1$ is maximal), demonstrating that $P_0\approx P_2$, indicating that $\mu\approx U/2$. (b) $\mu$ and $T$ (blue and red, respectively) are extracted from $P_0...P_3$ and the partition function, and plotted as a function of the repumping rate R. It is apparent that the lowest temperature occurs at the optimal repumping rate, and that the chemical potential is $\mu\approx U/2$ at this optimal operating configuration.}
\end{figure}

The partition function of a single qubit with anharmonicity $U$ in contact with a thermal bath at temperature $T$ is:
\begin{equation}
Z=\sum_{n=0}^{\infty}{e^{-\left[n(n-1)U/2-n\mu\right]/{k_B T}}}
\end{equation}

where $\mu$ is the chemical potential of the bath. In the main text, we engineer the thermalizer to maintain near-unity occupancy ($Pr(n=1)\approx 1$) with equal contributions of doubles and holes ($Pr(n=0)\approx Pr(n=2)\ll 1$), so, as shown in Figs. \ref{fig:SIchempotl} \& \ref{fig:SIchempotlNUMs}, we operate near $\mu\sim U/2$, at which point the only other terms that contribute substantially to $Z$ are $n=0$ and $n=2$, and we can write:
\begin{equation}
Z\approx 1+e^{U/{2k_B T}}+1
\end{equation}

We then compute $P_1$, the probability of a single photon excitation of the qubit, according to:

\begin{equation}
P_1\equiv Pr(n=1)= e^{U/{2k_B T}}/Z\approx 1-2 e^{-U/{2k_B T}}
\end{equation}

Employing $\left<1-P_1\right>_{optimal}\approx2\sqrt{\frac{6\Gamma_1}{U}}$ from the main text (for the two-site thermalizer), we can then solve for the temperature, and arrive at $\frac{k_B T}{U}\approx \frac{1}{\log{\frac{U}{24\Gamma_1}}}$.

To compute the entropy of the thermalizer qubit we employ $S\equiv-k_B\sum_N{Pr(n=N)\log{Pr(n=N)}}$, and arrive at the result (again for $\mu=U/2$ and near-unit occupancy):

\begin{align}
S&\approx F \frac{e^F}{2+e^F}-\log\left({2+e^F}\right)\nonumber\\
F&\equiv \frac{U}{2k_B T}
\end{align}

For $S\ll1$ this expression may be approximated by $\frac{S}{k_B}\approx 2 e^{-\frac{U}{2k_B T}}\left(1+\frac{U}{2 k_B T}\right)\approx\sqrt\frac{24\Gamma_1}{U}$.

\section{Incompressibility: Connecting $\frac{\partial V}{\partial P}$ with the gap between particle- and hole- bands}
\label{sec:SIcompressibility}

Following footnote 22 on page 708 of \cite{fazekas1999lecture}, pages 9-10 of~\cite{macdonald1994introduction}, and related ideas in \cite{kraemer2003bose}: if we call the volume of a system  $V$, the pressure $P$, and the total energy $U$, the compressibility is defined by $\kappa^{-1}\equiv-V\frac{\partial P}{\partial V}=V\frac{\partial^2 U}{\partial V^2}$.

The key point is that if the local density approximation applies (as it almost always does, even for relatively inhomogeneous systems ~\cite{Bakr2010}), then the energy per particle $u$ depends only on the particle density $\rho\equiv N/V$ according to $U=N u\!\!\left[\frac{N}{V}\right]$, we can rewrite $\kappa^{-1}=\rho^2 \frac{\partial \mu}{\partial \rho}$, where the chemical potential $\mu=\frac{\partial U}{\partial N}$.

Finally, we can write: $\kappa^{-1}=V \rho^2 \frac{\partial^2 U}{\partial N^2}$. This last expression may be interpreted to mean that if there is a discrete step in the energy cost to add a particle, then the inverse compressibility is infinite, so the compressibility is zero. A discrete step in the energy cost to add a particle is equivalent to a finite difference between the cost to add a particle and remove a particle- a spectral gap between particle and hole bands! This completes the connection.

\section{Repumping Doublons}
\label{sec:SIrepumpingdoublons}
Because the thermalizer's $U$ is assumed to be the same as the $U$ of the Bose-Hubbard chain,  the thermalizer is capable of refilling (really, evaporating) particle-defects (which in a Mott insulator take the form of doublons \cite{cheneau2012light}); any particle that hops into the already-populated thermalizer will be immediately evaporated via its resonator-enhanced loss process $\Gamma_2$.

Because $\Gamma_2=4R$ under optimal conditions, doublon-tunneling into the thermalizer will be Zeno-suppressed. To evaporate doublons efficiently, it is thus  favorable to include an additional lossy resonator, energetically tuned to $U$ and coupled directly to the Bose-Hubbard chain. Noting that doublons tunnel with a rate of $\sqrt{2}J$ and following the logic of Sec. \ref{sec:MBCoupling}, it is optimal to couple a lossy resonator to the other end of the chain with a strength $\sqrt{2}\times\sqrt{2}J=2J$, and, identifying the lossy resonator linewidth with the ``doublon repumping rate,'' a loss-resonator linewidth properly matched to the observed doublon production rate.

\bibliography{ThermalizationBib}{}

\end{document}